\documentclass[aps]{revtex4}
\usepackage{amssymb,amsmath,bm,graphicx,color,subfigure}
\begin{document}
\title{Dynamics of confined L\'{e}vy  flights in terms of (L\'{e}vy) semigroups}
\author{Piotr Garbaczewski  and  Vladimir Stephanovich}
\affiliation{Institute of Physics, University of Opole, 45-052 Opole, Poland}
\date{\today }
\begin{abstract}
  The master equation for a probability density function (pdf) driven by   L\'{e}vy  noise,
   if conditioned to  conform with   the principle of  detailed balance,
   admits a transformation to a contractive strongly continuous semigroup  dynamics.
   Given a priori a  functional form of the  semigroup potential,  we address the ground-state  reconstruction problem for
    generic  L\'{e}vy-stable  semigroups,   for {\em all} values of the stability index $\mu  \in (0,2)$.
   That is  known to  resolve  an  invariant  pdf  for  confined L\'{e}vy flights (e.g.  the  former
     jump-type process).  Jeopardies of the procedure are discussed, with a  focus  on:  (i) when an invariant  pdf actually
     is an asymptotic one,  (ii) subtleties of the   pdf    $\mu $-dependence in the vicinity   and  sharply {\em at}   the
      boundaries $0$ and $2$ of the  stability interval,     where jump-type scenarios  cease to be valid.
\end{abstract}
\maketitle

\section{Conceptual background}

As  the   L\'{e}vy  noise model we   consider  any member of   a subclass of  uni-variate stable probability distributions  determined by
a characteristic exponent  $ -F(p) =  -|p|^{\mu }$ of $<\exp(ipX)>$,  with   $0< \mu < 2$.  The induced
 free  jump-type   dynamics,  $<\exp(ipX_t)>= \exp[-t F(p)]$, where $t\geq 0$, is  conventionally interpreted in terms of L\'{e}vy flights
  and  quantified  by means of a pseudo-differential (fractional) equation for a  corresponding  time-dependent probability density function (pdf)
\begin{equation} \label{master}
\partial _t \rho  = -  |\Delta |^{\mu /2} \rho =  \int [w_{\mu }(x|y) \rho (y) - w_{\mu }(y|x) \rho (x)]dy\, .
\end{equation}
  The jump rate  $ w_{\mu } (x|y)\propto 1/|x-y|^{1+\mu }$ is  a symmetric  function,
$w_{\mu }(x|y)= w_{\mu } (y|x)$.

 The  "free"  fractional Fokker-Plack equation   (\ref{master}) has  no stationary solutions.  On  the other hand,
asymptotic invariant pdfs   for confined  L\'{e}vy flights are  known to arise in the standard Langevin modeling  of
 an  external forces impact on the  L\'{e}vy-stable noise, \cite{klafter0}.  A disadvantage of that  approach is
  a non-existence of Boltzmann-type (thermal) equilibria, \cite{klafter}-\cite{sokolov}. We note  in passing
  that the reference stable laws  generically  have no  moments of order higher than one (and may as well have none at all). To the contrary,
     pdfs  for confined L\'{e}vy flights may have an arbitrary,  not necessarily finite  number of moments.

Interestingly, if we  enforce  \cite{gar0,sokolov}  the principle of detailed balance  to hold true,   by a   suitable  modification of transition rates
  $w(x|y)\rightarrow w_U(x|y)$ (with $U$ playing the role of   an external microscopic  potential),
   asymptotic  Boltzmann-type equilibria of the form $\rho (x) \sim \exp[-U(x)]$ in principle  become admissible.
      The price paid is that the  standard  Langevin modeling  of confined L\'{e}vy flights  becomes  inadequate.

We generalize the master equation (\ref{master})  to encompass non-symmetric    jump rates
 as follows,   $ w_{\mu}(x|y)  \rightarrow  w_{\mu }^U(x|y) \neq w_{\mu }^U(y|x)$:
\begin{equation}
w_{\mu}^U(x|y) = w_{\mu }(x|y) \, \exp \left(\frac{U(y) - U(x)}{2}\right),
\end{equation}
where $U(x)$ is a continuous   function on $R$.
With $w_{\mu }^U(x|y)$ replacing $w_{\mu }(x|y)$,  Eq.  (\ref{master})   takes the  form
\begin{equation}\label{kinetic}
\partial _t \rho =
-  [\exp (-U/2)]\, |\Delta |^{\mu /2}[ \exp(U/2 )
    \rho ]   +    \rho \exp (U/2 ) |\Delta |^{\mu /2} \exp(-U/2)\, .
\end{equation}
For a suitable (to secure normalization)  choice of  $U(x)$, $\rho _{eq}(x) \propto \exp [- U(x)] $ is a stationary solution of Eq.~(\ref{kinetic}).
 The  detailed balance principle necessarily  follows:
\begin{equation}
 w_U(x|y)\rho _{eq}(y)=  w_U(y|x)\rho _{eq}(x)\, .
\end{equation}

 The master equation (\ref{kinetic})  cannot be derived within the multiplicative or additive Langevin modeling.

 As  mentioned before,  Eq.~(\ref{kinetic}) at least on formal grounds,  admits a transformation to a strongly continuous semigroup dynamics.
   That is akin to a mapping of the standard  Fokker-Planck equation into the generalized diffusion (semigroup) equation  which
     is   widely exploited in  the context of  diffusion-type processes  (there e.g.  the Fokker-Planck operator  is mapped into
      a symmetric operator,  which furthermore  needs to be extended to a self-adjoint one).
  In the  Brownian case, the  Langevin and semigroup derivations, while interpreted in terms of $\rho (x,t)$,   refer to the
   same diffusion-type process. To the contrary,   this  equivalence {\it  does  not} persist  in case of confined L\'{e}vy flights.

The passage from Eq.~(\ref{kinetic}) for confined L\'{e}vy flights  to the semigroup dynamics  is accomplished
by means of a redefinition
\begin{equation}
\rho (x,t) = \rho _*^{1/2} (x) \Psi (x,t)
\end{equation}
where  $\rho _*(x) = \rho _{eq}(x)= Z^{-1} \exp [- U(x)] $ is an asymptotic invariant pdf of the jump-type process,
while the dynamics of  a {\it real} positive-definite function  $\Psi (x,t)$ follows   the semigroup pattern $[\exp(-\hat{H}t)\,\Psi ](x,0) = \Psi (x,t)$
for $t\geq 0$, c.f. \cite{sokolov,gar1,stef1}.  Here, we have  introduced   the L\'{e}vy-Schr\"{o}dinger Hamiltonian operator  with an  external potential
\begin{equation}\label{hamiltonian}
\hat{H}_{\mu }  \equiv   |\Delta |^{\mu /2} +  {\cal{V}}(x)\, .
\end{equation}
Suitable  properties of ${\cal{V}}$ need to be assumed, so that
 (i) $-\hat{H}_{\mu }$ is a legitimate (self-adjoint)  generator  of a   (strongly continuous, contractive)
  semigroup  $\exp(-t \hat{H}_{\mu })$, (ii) asymptotically, as $t\rightarrow \infty $ we get
      $\Psi (x,t)\rightarrow \Psi _*(x) \sim \rho ^{1/2}_*(x)$ and  $ \Psi _*(x)$ is a unique ground state of $\hat{H}$.
Said otherwise, once we have  a priori selected an invariant probability density  $\rho _{eq}(x) \doteq \rho _*(x)
\propto \exp[-U(x)]$,  (\ref{kinetic}) to justify its interpretation as  an asymptotic pdf of a well
 defined  jump-type  process    we must  have guaranteed  an  existence of an associated contractive
  semigroup dynamics for which $\Psi _*(x) \sim \rho ^{1/2}_*(x)$.

Looking for stationary solutions of the  semigroup  equation
$\partial _t \Psi = - \hat{H}_{\mu } \Psi$, we realize that if a square root of a  positive  invariant   pdf $\rho _*(x)$  is asymptotically  to  come out via  the dynamics  $\Psi \rightarrow  \rho _*^{1/2}$, then the resulting
fractional Sturm-Liouville equation $\hat{H}_{\mu }  \rho _*^{1/2}=0$  stands for  a {\em{compatibility condition}}
 upon the functional form of  ${\cal{V}}(x)$:
 \begin{equation}\label{comp}
 {\cal{V}}  =   -\frac{|\Delta |^{\mu /2}  \rho ^{1/2}_*}{\rho ^{1/2}_*}.
 \end{equation}
We note that the inferred  semigroup dynamics provides a solution for  the L\'{e}vy stable {\em {targeting problem}}, with a predefined invariant pdf.
It is an identification  $\rho (x,t) =  \rho _*^{1/2}  (x)\Psi (x,t)$  that  does   the job.
 We have discussed this issue in some detail in our previous publications, \cite{gar,gar1,stef1}.

Inversely,  if we  choose a priori a  concrete potential function ${\cal{V}}(x)$, then  an ultimate   functional form of an
 invariant  pdf  $\rho _*(x)$ (actually $\rho _*^{1/2}(x)$), if in existence,     needs to  come out from the  above  compatibility condition.
 However, a solvability of \eqref{comp} with respect to $\rho _*^{1/2}$ merely  associates an {\it invariant} pdf  $\rho _*$   with a pre-defined semigroup potential
 ${\cal{V}}$.  In  regard to the  {\it  asymptotic regime}  $\Psi (x,t) \rightarrow \rho ^{1/2}_*$ we need much more.  Namely, $\rho ^{1/2}_*(x)$
  must belong to the domain of  a self-adjoint operator $\hat{H}$  and a proper identification of such domain may sometimes become tricky.

  The latter  problem, and jeopardies involved,  we  address in the present paper for  a  predefined function    ${\cal{V}}(x)$, while admitting  all
  stability index values  $0<\mu <2$ in the compatibility condition (\ref{comp}).

{\bf Remark 1:} For clarity of presentation and self-explanatory features  of  discussion  we add few comments about the stochastic process in question.
 Let $\hat{H}$ be a self-adjoint operator in a suitable Hilbert space domain. Additionally let  ${\cal{V}}={\cal{V}}(x)$ be a  bounded
from below  continuous function. Then,  the  integral kernel $k(y,s,x,t)=\{ \exp[-(t-s)\hat{H}] \} (y,x)$, $s<t$, of the
semigroup  operator  $\exp(-t\hat{H})$  is  positive and jointly continuous in all variables.   The semigroup dynamics reads:
$\Psi  (x,t) = \int \Psi (y,s)\, k(y,s,x,t)\,  dy$ so that for all $0\leq s<t$  we can reproduce the dynamical pattern of  behavior,
  actually  set by  Eq.~(\ref{kinetic}), but now in terms of Markovian  transition probability densities $p(x,s,y,t)$:
$\rho (x,t) =  \rho _*^{1/2}  (x)\Psi (x,t) = \int p(y,s,x,t) \rho (y,s) dy$,
where  $p(y,s,x,t) = k(y,s,x,t)\, \rho _*^{1/2}(x)/ \rho _*^{1/2}(y).$
An asymptotic behavior of $\Psi (x,t)\to \rho _*^{1/2}(x)$ implies  $\rho (x,t) \rightarrow \rho _*(x)$  as $t\rightarrow \infty $.

{\bf Remark   2:}  The spectral theory of fractional operators of the form (\ref{hamiltonian}) has received a broad coverage in the mathematical
 \cite{davies,bertoin,kaleta,lorinczi,lorinczi1} and mathematical physics  literature \cite{carmona,carmona1}.
  Various rigorous estimates pertaining to the decay of the eigenfunctions at spatial   infinities, quantify
  the number of moments of the associated  pdfs for  different  classes of potential functions ${\cal{V}}(x)$.
  As well, fractional versions of the Feynman-Kac formula for  an integral kernel of the semigroup  operator have an ample coverage therein.

\section{$\mu $-family of semigroup potentials for a predefined  invariant pdf and the  $\mu \in (0,2)$ boundary issue}
For a pseudo-differential operator $|\Delta |^{\mu /2}$,
    the action on a function from  its domain is  greatly simplified in the physics-oriented research.  Normally,  with $\nu _{\mu }(dx)$
   standing for the L\'{e}vy measure, we have:
\begin{equation}(|\Delta |^{\mu /2} f)(x)\, =\, - \int_R [f(x+y) - f(x) - {{y\, \nabla f(x)}
\over {1+y^2}}]\, \nu _{\mu }(dy)
\end{equation}
By turning over to the Cauchy principal value  of the involved  integral, one  actually considers
\begin{equation}
 (|\Delta |^{\mu /2} f)(x)\, =\, - \int  [f(x+y) - f(x) ] \nu _{\mu }(dy) \,
\end{equation}
which upon changing  an integration variable $y\rightarrow z=x+y$, may be reproduced in  form, see e.g. \cite{sokolov}
\begin{equation}\label{def}
 -(|\Delta |^{\mu /2} f)(x)\,  =
   {\frac{\Gamma (\mu +1) \sin(\pi \mu/2)}{\pi }} \int  {\frac{f(z)- f(x)}{|z-x|^{1+\mu }}}\,
 dz \,\, .
\end{equation}

Let us investigate the properties of the   $-|\Delta|^{\mu/2} f(x)$ by turning over to the   Fourier image of $f(x)$.
We  employ   a redefinition of Eq.~(\ref{def})
\begin{equation}\label{mut0}
    -|\Delta|^{\mu/2}f(x)=\frac{\Gamma(1+\mu)\sin\frac{\pi \mu}{2}}{\pi}\int_{-\infty}^{\infty}dy\frac{f(x+y)-f(x)}{|y|^{1+\mu}}.
\end{equation}
which yields
\begin{eqnarray}
-|\Delta|^{\mu/2}f(x)=\frac{\Gamma(1+\mu)\sin\frac{\pi \mu}{2}}{\pi \sqrt{2\pi}}
\int_{-\infty}^{\infty}f(k)e^{-\imath kx}dk \int_{-\infty}^{\infty}\frac{(e^{-\imath ky}-1)dy}{|y|^{1+\mu}} .\label{intf1}
\end{eqnarray}
The integral over $dy$ can be calculated as follows
\begin{equation}\label{intf2}
 \int_{-\infty}^{\infty}\frac{(e^{-\imath ky}-1)dy}{|y|^{1+\mu}}\equiv 2 \int_{0}^{\infty}\frac{(\cos ky -1)dy}{|y|^{1+\mu}}=
 2|k|^\mu\Gamma(-\mu)\cos\frac{\pi \mu}{2}.
\end{equation}
It is seen that at  the limiting value  $\mu=0$,  $\Gamma(0)$ is divergent,  so that  the integral \eqref{intf2} is divergent as well.  The same happens for $\mu =2$, in view of the divergence of    $\Gamma(-2)$.
 However, irrespective of how close to  $0$  or  $2$  the label $\mu >0$ is,   the  integral \eqref{intf2} is convergent.

 It is  interesting to observe that  the   divergence of the  the Fourier integral, as $\mu $ approaches $0$ or $2$,  becomes compensated,
  if we substitute  it back to Eq. \eqref{intf1} and next consider  the limiting behavior of the result:
\begin{eqnarray}\label{intf3}
&&-|\Delta|^{\mu/2}f(x)=\frac{2\Gamma(1+\mu)\Gamma(-\mu)\sin\frac{\pi \mu}{2}\cos\frac{\pi \mu}{2}}{\pi \sqrt{2\pi}}
 \int_{-\infty}^{\infty} |k|^\mu f(k) e^{-\imath kx}dk=\nonumber \\
 &&=- \frac{1}{\sqrt{2\pi}}\int_{-\infty}^{\infty} |k|^\mu f(k)
  e^{-\imath kx}dk.
 \end{eqnarray}
  Here we use the identity
 \begin{equation}\label{mag}
 \Gamma(1+\mu)\Gamma(-\mu)=-\frac{\pi}{\sin \pi \mu}.
 \end{equation}

In view of the above divergence obstacle, the integral representation (\ref{def}) is  invalid  \it at \rm  the boundaries of the stability interval. However, the range of validity of  its ultimate
 Fourier version, e.g. the right-hand-side of  Eq.~(\ref{intf3}),  can be safely  extended to the   boundary values $0$ and $2$.

  In fact, it is $ - {\frac{1}{\sqrt{2\pi}}} \int_{-\infty}^{\infty} |k|^\mu f(k)
  e^{-\imath kx}dk  \equiv   - \partial _{\mu }/\partial |x|^{\mu }  \equiv (-\Delta )^{\mu /2}$   that is commonly interpreted  in the literature
  as a definition of the fractional derivative  of the $\mu $-th order. On formal grounds  this definition encompasses the boundary cases $(-\Delta )^0\equiv 1$ and  $ -\Delta$.

\subsection{Heavy-tailed case}
Let us first consider a specific pdf (member of the so-called Cauchy family, \cite{gar})
$\rho (x)= 2/[\pi (1+x^2)^2]$, such that   $\rho ^{1/2}(x)=\sqrt{\frac{2}{\pi}}{\frac{1}{1+x^2}}$ and its Fourier transform reads
$\rho ^{1/2}(k)=e^{-|k|}$.  In view of $\int_{-\infty}^{\infty}e^{-\imath kx} f(|k|)dk\equiv 2 \int_{0}^{\infty}\cos(kx) f(k)dk$  we have
 \begin{eqnarray}\label{mutar7}
{\cal V}_{\mu }(x)=-(1+x^2)\int_{0}^{\infty} k^{\mu}e^{-k}\cos kx\ dk=-(1+x^2)^{\frac{1-\mu}{2}}\Gamma(1+\mu)
\cos\left[(1+\mu)\arctan x\right],\ 0<\mu<2.
\end{eqnarray}
Expression \eqref{mutar7} permits to reproduce easily a number of ${\cal V}_{\mu }$ for different values of $\mu \in (0,2)$.
All  ${\cal V}_{\mu }$, $\mu \in (0,2)$,   derive from a  pre-defined (quadratic Cauchy)   pdf.
   However, we find most interesting the  following three  limiting cases.
Namely,   for $\mu =1$, we get the  result derived previously in Ref.~\cite{gar1}
\begin{equation}\label{mutar10}
   {\cal V}_1(x)= \frac{x^2-1}{x^2+1}.
\end{equation}
In the vicinity of the boundary value $\mu =2$
the expression \eqref{mutar7}   yields
\begin{widetext}
\begin{equation}\label{mutar12}
   {\cal V}_{\mu \to 2,2}(x)\approx \frac{2(3x^2-1)}{(1+x^2)^2} -\frac{\mu-2}{(1+x^2)^2}\left[2x(x^2-3)\arctan x
   +(3x^2-1)\left(2\gamma -3+\ln(1+x^2)\right)\right],
\end{equation}
\end{widetext}
where $\gamma \approx 0.577216$ is Euler constant.
Hence, for  the boundary value $\mu = 2$  we get
\begin{equation}\label{mutar9}
   {\cal V}_{2}(x)= \frac{2(3x^2-1)}{(1+x^2)^2}.
\end{equation}
which stays in conformity with a naive expectation that  $ -|\Delta | \equiv \Delta $.
Indeed, of formal grounds,  we  readily get \eqref{mutar9} via  $
    {\cal V}_{2}(x) \rho ^{1/2}(x)  =+{d^2}{ \rho ^{1/2}(x)}/{dx^2} $, as expected.

For $\mu =0$, in view of   $\arctan x= \arccos (1/\sqrt{1+x^2})$, from (12)  we recover ${\cal V}_{0}(x)=-1$ which is   compatible with  $ {\cal V}_{0}(x)=  - {\frac{|\Delta |^0\rho ^{1/2}_*}{\rho ^{1/2}_*}}=-1$
upon an  identification $ |\Delta |^0  \equiv 1$.

\subsection{Gaussian pdf}

Let us consider   the invariant  pdf in a Gaussian form  $
\rho_*=\frac{1}{\sigma\sqrt{2\pi}}e^{-\frac{x^2}{2\sigma^2}}$ whose  square root  $\rho _*^{1/2}$
has the Fourier image
\begin{equation}\label{dyb3}
(\rho^*)^{1/2}(k)\equiv f(k)=\frac{1}{\sqrt{\sigma}\sqrt[4]{2\pi}}\frac{1}{\sqrt{2\pi}} \int_{-\infty}^{\infty}
e^{-\frac{x^2}{4\sigma^2}}e^{ikx}dx=
\sqrt{\sigma}\sqrt[4]{\frac{2}{\pi}}e^{-k^2\sigma^2}.
\end{equation}
Accordingly
\begin{eqnarray}\label{dyb4}
{\cal V}_{\mu }(x)&=&-\frac{1}{(\rho^*)^{1/2}(x)}\frac{1}{\sqrt{2\pi}} \int_{-\infty}^{\infty}(\rho^*)^{1/2}(k)|k|^\mu
e^{-ikx}dk=\nonumber\\
&=&-\frac{2\sqrt{\sigma}}{(\rho^*)^{1/2}(x)\sqrt[4]{2\pi}\sqrt{\pi}}\int_{0}^{\infty}|k|^\mu \cos kx \ e^{-k^2\sigma^2}dk.
\end{eqnarray}
Clearly, at $\mu  = 2$ we arrive at
 \begin{equation}\label{dyb5}
{\cal V}_{2}(x)=\frac{1}{2\sigma^2}\left(\frac{x^2}{2\sigma^2}-1\right),
\end{equation}
whose equivalent derivation is provided by setting   $ -|\Delta | \equiv +\Delta $ in Eq. (\ref{comp}).  Indeed, we verify by inspection that
${\cal V}_{2}(x)= \frac{f''(x)}{f(x)}$.

The case of $\mu = 0$ can be handled as before with the outcomes  ${\cal V}_{0}(x)=-1$ and   $ |\Delta |^0  \equiv 1$.
To this end we observe that (set $\mu =0$ in Eq.~(13))
$\int_{0}^{\infty}\cos kx \ e^{-k^2\sigma^2}dk={\frac{\sqrt{\pi}}{2\sigma}}  \exp ({-\frac{x^2}{4\sigma^2}})$ while
$(\rho^*)^{1/2}={\frac{\sqrt{\sigma}}{\sqrt[4]{2\pi}}}  \exp ({-\frac{x^2}{4\sigma^2}})$.

\section{$\mu $-family of  pdfs  for a predefined ${\cal{V}}(x)$: generalties.}

Presently,  we shall proceed in reverse. If we choose  a priori a  concrete potential function ${\cal{V}}(x)$, then  an ultimate   functional form of an
 invariant  pdf  $\rho _*(x)$ (actually $\rho _*^{1/2}(x)$),  \it  if in existence\rm,      needs to  come out from the   compatibility condition  \eqref{comp}.
For any functional form of   the confining potential ${\cal{V}}(x)$,  Eq.~\eqref{comp} imposes the following  identity to be obeyed by
 an invariant (terminal) pdf (we denote $\rho
^{1/2}_*(x)\equiv f(x)$)
\begin{equation}\label{bu1}
     {\cal V}(x)f(x) = - |\Delta |^{\mu/2}f(x),
\end{equation}
where $0<\mu<2$.
Remembering that we  consider ${\cal{V}}(x)$ to be a continuous and bounded from below function (may be unbounded from above),
we turn over to  the standard Fourier transform method with $ f(k)=\frac{1}{\sqrt{2\pi}}\int_{-\infty}^{\infty}f(x)e^{\imath kx}dx$ and
$f(x)=\frac{1}{\sqrt{2\pi}}\int_{-\infty}^{\infty}f(k)e^{-\imath kx}dk$.
Denoting  $u_k$ the  Fourier image  of the right-hand side of Eq. \eqref{bu1}, we obtain
\begin{equation}\label{bu1a}
u_k=-|k|^\mu f(k).
\end{equation}
Equating  Fourier images of both sides of Eq. \eqref{bu1}, provided
they exist,  yields
\begin{equation}\label{bu2}
u_k=\frac{1}{\sqrt{2\pi}}\int_{-\infty}^{\infty}{\cal V}(x)f(x)e^{\imath kx}dx=
\frac{1}{2\pi}\iint_{-\infty}^{\infty}{\cal V}(x)e^{\imath x(k-k')}f(k')dk'dx.
\end{equation}
In this case, the Fourier image $f(k)$  of a solution  $f(x)$ to Eq. \eqref{bu1} is defined by following
integral equation of a convolution type
\begin{equation}\label{bu5}
    f(k)=-\frac{1}{|k|^{\mu}\sqrt{2\pi}} \int_{-\infty}^{\infty} {\cal V}(k-k')f(k')dk'.
\end{equation}
Here we employ  the identity $
\frac{1}{2\pi}\int_{-\infty}^{\infty}e^{\imath
x(k-k')}dx=\delta(k-k')$,    valid in the sense of distributions.

Our confining assumption implies that the  function ${\cal
V}_\mu(x)$   typically   grows  at infinities so that its  Fourier
image may not exist, unless distributionally.  The same jeopardy
appears in case of a product ${\cal{V}}(x)f(x)$, (\ref{bu2}). Modulo
those obstacles, we shall demonstrate  that, if interpreted  in the
sense of distributions,  the general solution of Eq. \eqref{bu5} can
be expressed in terms of $\delta$-function and its derivatives.

Let us restrict further consideration to even   functions
${\cal{V}}$ (in this case the terminal pdf is an even function as
well) and  for clarity of presentation  assume them
 to be  entire functions.
 The Taylor series comprise even powers of $x$ only
\begin{equation}\label{yta1}
    {\cal V}(x)={\cal V}(0) + {\cal V}"_\mu(0)\frac{x^2}{2!}+ {\cal V}^{(4)}_\mu(0)\frac{x^4}{4!}+...
\end{equation}
The Fourier image of \eqref{yta1},  c.f. \eqref{bu2}  and
\eqref{bu5}, yields
\begin{eqnarray}
&&{\cal V}(k)=\frac{1}{\sqrt{2\pi}}\int_{-\infty}^{\infty}\left[{\cal V}(0) +
 {\cal V}"(0)\frac{x^2}{2!}+ {\cal V}^{(4)}(0)\frac{x^4}{4!}\right] e^{\imath kx}dx \equiv \nonumber \\
&&\equiv \sqrt{2\pi}\left[{\cal V}(0)\delta(k)-\frac{{\cal V}"(0)}{2}\delta"(k)+\frac{{\cal V}^{(4)}(0)}{4!}\delta^{(4)}(k)-...\right],
\label{yta2}
\end{eqnarray}
i.e. it has the form of the infinite series of even derivatives of the Dirac  $\delta$ - function. We note that if ${\cal V}(x)$
is a  simple (even) polynomial (example:  ${\cal V}(x)=ax^2+bx^4$), the above series are finite.

Accordingly, we end up with the following differential equation of the infinite even order for the Fourier image  $f(-k)=f(k)$ of $\rho^{1/2}_*(x)\equiv f(x)$:
\begin{equation}\label{yta5}
\frac{{\cal V}"(0)}{2}\frac{d^2f(k)}{dk^2}- \frac{{\cal V}^{(4)}(0)}{4!}\frac{d^4f(k)}{dk^4}+...=\biggl[k^\mu+{\cal V}(0)\biggr]f(k),\ k\geq 0.
\end{equation}
We choose the following initial conditions  for  Eq.~(\ref{yta5})
\begin{equation}\label{yta6}
f(k=0) \equiv \int_{-\infty}^{\infty}f(x)dx=A,\ f^{(2n-1)}(k=0)=0,\ n=1,2,3,....
\end{equation}
Note that this imposes  an integrability condition on  $\rho _*^{1/2}(x)$ on ${\cal R}$.
Here by $f^{(2n-1)}(k=0)$ we denote the odd derivatives of $f(k)$ at $k=0$.

The integration  constant $A$  is not completely  arbitrary and should   be consistent with the normalization condition
 $\int_{-\infty}^{\infty}f^2(x)dx \equiv  \int_{-\infty}^{\infty}\rho_*(x)dx=1$.  In view of the Parceval identity  we have
\begin{equation}\label{yta8}\int_{-\infty}^{\infty}f^2(x)dx=\int_{-\infty}^{\infty}f^2(k)dk \equiv 2 \int_{0}^{\infty}f^2(k)dk=1.
\end{equation}
This means that $f(k=0)=A$  must be compatible  with  $\int_{0}^{\infty}f^2(k)dk=1/2$.

One should not expect an easy analytic outcome of the solution of the infinite order differential equation \eqref{yta5}.
In most cases of interest the infinite series can be truncated, but generically  a numerical assistance is unavoidable.
The practical strategy of finding (at worst approximately for truncated series and an  arbitrary functional  shape of ${\cal V}(x)$) an $L^2({\cal R})$ integrable non-negative ground state of the  a  priori prescribed semigroup, and thence the terminal pdf $\rho_*(x)$,  can be summarized as follows.
\begin{itemize}
\item Expand ${\cal V}(x)$ in power series. The number of terms in the series should be chosen so as to obtain
a sufficiently good approximation of the potential.
\item Solve  the differential equation \eqref{yta5}  with initial conditions \eqref{yta6}. If ${\cal V}(x)$
is a polynomial function, there are good chances to  solve this equation analytically. Otherwise we should reiterate to numerics.
Check a compatibility of  $f(k=0)=A$ with the  normalization condition.
\item Analytically or numerically  take the inverse Fourier transform to obtain  a non-negative
 function  $f(x)$, to be interpreted as $\rho^{1/2}_*(x)$.
\item Check whether the obtained  $f(x)$, $x^2f(x)$  and  $|\Delta |^{\mu /2}f(x)$ are absolutely integrable (to guarantee that they actually
 are Fourier transformable). Additionally check that they  are  square integrable functions  and $L^(R)$ normalize the resultant $f(x)$.
\item  Everything  is being  accomplished under an  assumption that $ |\Delta |^{\mu /2} +  {\cal{V}}(x)$  is not merely symmetric, but
 a self-adjoint operator in  a suitable domain that is dense in $L^2(R)$. A tacit assumption is that  $\rho^{1/2}_*(x)$ actually
  {\it is} in the domain of $\hat{H}$. However, this point is quite delicate and involves jeopardies,   as we shall see in what follows.
  The operators considered are always symmetric. An issue of their self-adjoint extensions is by no means trivial.
  Only  a self-adjoint generator  allows to interpret  the  derived  pdf   as  the   terminal  (asymptotic invariant)  one for an associated
    semigroup dynamics
\end{itemize}

\section{Case study of  the pdf  reconstruction   for    ${\cal{V}}=x^2/2$ and   $\mu  \in (0,2)$}

We shall consider  an exemplary  analytic  (in part computer-assisted)  realization of the previously  outlined    inverse procedure.
Our  motivations come from Ref.~\cite{gar1}, where the Cauchy oscillator has been addressed.

 We recall that  by analogy with the
familiar harmonic oscillator Hamiltonian (related to the Ornstein-Uhlenbeck process)
$\hat{H} \equiv - D \Delta   + \left( {\frac{\gamma ^2 x^2}{4D}}  - {\frac{\gamma }2} \right)$, whose non-negative   spectrum  starts from $0$,
  we have considered,   \cite{gar1,lorinczi1}, the Cauchy oscillator problem in the form
$\hat{H}_{1/2 }  \equiv  \lambda |\nabla |  + \left( {\frac{\kappa }2} \,  x^2  - {\cal{E}}_0\right)$, with  ${\cal{E}}_0$  left unspecified.
The compatibility condition (6) appears in the form $\left({\frac{\kappa }2} \,  x^2  - {\cal{E}}_0\right) \rho ^{1/2}_* =   -\lambda\,  |\nabla |\,
 \rho ^{1/2}_*$. Its   Fourier transform  reads
\begin{equation}\label{rai1}
-{\frac{\kappa }2} \Delta _p \tilde{f} +  \gamma |p| \tilde{f}= {\cal{E}}_0 \tilde{f}
\end{equation}
where $\tilde{f}(p)$  stands for the Fourier transform of $f=\rho _*^{1/2}(x)$.

By changing an independent variable $p$ to $k= (p-\sigma)/\zeta$, next denoting $\psi (k) = \tilde{f}(p)$ with the  identifications
$\sigma = {\cal{E}}_0/\gamma $ and $\zeta = (4\kappa /\gamma )^{1/3}$, we may rewrite the above eigenvalue problem (with ${\cal{E}}_0$
standing for an eigenvalue) in the form of the following ordinary differential  equation
\begin{equation} \label{rai2}
{\frac{d^2 \psi (k)}{dk^2}} = 2 |k| \psi (k),
\end{equation}
whose solutions can be represented in terms of Airy functions, c.f. Ref.~\cite{gar1}.  We note in passing that a slightly
 different scaling was  originally  employed in Ref.~\cite{gar1}: $\zeta = (\kappa /2\gamma )^{1/3}$ $\rightarrow $
 ${\frac{d^2 \psi (k)}{dk^2}} =  |k| \psi (k)$.

  The  equation  (\ref{rai2})   is  in fact  a departure point for our subsequent discussion.
Quite at variance with the standard  (text-book) harmonic oscillator  intuitions and our own discussion of
the Cauchy oscillator spectral problem \cite{gar1}, just out of curiosity and as a useful exercise
 let us  consider  the compatibility condition
for  ${\cal{V}}=x^2/2$  proper, e.g.
{\it  without} any additive counterterms of the form $-{\cal{E}}_0$:
\begin{equation}\label{terp1}
  \frac{x^2}{2}\rho ^{1/2}_*= - |\Delta |^{\mu/2}\rho ^{1/2}_*, \ 0<\mu<2.
\end{equation}
That  amounts to assigning   the  "improper" eigenvalue zero as the bottom eigenvalue of the corresponding spectral problem.

We take Fourier images of  both sides of Eq.\eqref{terp1} to obtain
\begin{equation}\label{terp2}
u_k=\frac{1}{\sqrt{2\pi}}\int_{-\infty}^{\infty}\frac{x^2}{2}f(x)e^{\imath kx}dx=-\frac 12 \frac{1}{\sqrt{2\pi}}
\frac{\partial^2}{\partial k^2}\int_{-\infty}^{\infty}f(x)e^{\imath kx}dx\equiv -\frac 12 \frac{\partial^2 f(k)}{\partial k^2}
\end{equation}
Accordingly, we have
\begin{equation}\label{terp3}
    \frac{d^2 f(k)}{dk^2} =2|k|^\mu f(k)
\end{equation}
whose  special $\mu =1$ case  the previous Eq.~(\ref{rai2}) actually is.

To find a solution to Eq.~(\ref{terp3}), our main   idea here (and possibly for  an  arbitrary potential) stems from
 the  approach we  have  originally  adopted   for  Eqs.~(\ref{rai1}) and  (\ref{rai2}), \cite{gar1,rob}
 and  based on  the  exploitation of Airy functions.
 We note in passing  that both the    (Fourier) integral definition of the fractional operator and the solution  construction procedure involves
  the Cauchy principal value integral and the continuity  (gluing)  conditions  in the vicinity and ultimately at the point $0\in R$.

We should find the decaying solution of  the corresponding differential equation  Eq.~(\ref{terp3}) in the  $k$-space, on the  positive semi-axis ($k>0$),
and an  oscillatory one on the  negative semi-axis ($k<0$). Then we need to  shift the   obtained solution to the right so that the first maximum of
the oscillatory part is  located   at $k=0$. After "chopping" the rest of the  oscillating part  one has to   reflect the remaining  piece about
 the  vertical axis to get an   even "bell-shaped" function.

 The obtained    $k$-space   solution  should be Fourier-inverted
  (properly - here we encounter a  crucial  difference with the reasoning of \cite{gar1} for the Cauchy oscillator) and squared, while
  keeping in mind the $L^2(R)$ normalization issue.  This   procedure is expected to determine   the   desired   invariant
   (and a candidate for  terminal)  pdf in the $x$ -space.

{\bf Remark  3:}  We need to explain how our workings with the  $+k_m$   shift are  to be understood, since \eqref{terp3} itself
 is {\it not} translationally invariant.
  Eq. \eqref{terp3} has the form $d^2f/dk^2= G(k)f(k)$, where $G(k)$ can in principle be an arbitrary function.  We execute
$k \rightarrow   k+k_m$ everywhere in \eqref{terp3}, so   arriving at $d^2f/dk^2=G(k+k_m)f(k+k_m)$.
Obviously, with  $k'=k+k_m$  we have  $ d^2f/d{k'}^2=G(k')f(k')$, where  $k'\in R$ and the same as previously boundary data at infinities.
This function is actually  Fourier inverted with
respect to the $k'$-label.
 To see clearly, why such pre-caution needs to be kept in mind, lets us notice that $f(x)=\int_0^{\infty } f(k') cos(k'x)dk'$
 maps back  \eqref{terp3} into \eqref{terp1}.  On the other hand, to map \eqref{rai2} into the spectral problem for  $\hat{H}_{1/2 }  \equiv  \lambda |\nabla |
  + \left( {\frac{\kappa }2} \,  x^2  - {\cal{E}}_0\right)$,   we need to execute the inverse Fourier
 transform with respect  to   $p$ in $\psi(k) = \psi [k(p)]= \tilde{f}(p)$.

To solve  \eqref{terp3}  and next \eqref{terp1},  according to the previously outlined procedure,
let us  consider a pair of equations $\frac{d^2 f(k)}{dk^2} =2\ {\rm{sign}} k \ |k|^\mu f(k)$,  instead of  the single  \eqref{terp3}:
\begin{equation}\label{subs}
\left\{
\begin{array}{cc}
  \frac{d^2 f(k)}{dk^2} =2k^\mu f(k), & k>0  \\ \\
  \frac{d^2 f(k)}{dk^2} = -2(-k)^\mu f(k), & k<0.   \\
   \end{array}
\right.
\end{equation}
The resultant  solutions have  different forms  for  $k>0$ and $k<0$  respectively, \cite{polyanin}. Namely, for  $k \geq 0$  we have
\begin{equation}\label{terp4}
f(k)=\sqrt{k}\left[C_{11} I_{\frac{1}{2q}}\left(\frac{\sqrt{2}}{q}k^q\right)+
C_{12} K_{\frac{1}{2q}}\left(\frac{\sqrt{2}}{q}k^q\right)\right],\ q=\frac 12(\mu+2),
\end{equation}
while for  $k<0$ there holds
\begin{equation}\label{terp4m}
f(k)=\sqrt{|k|}\left[C_{21} J_{\frac{1}{2q}}\left(\frac{\sqrt{2}}{q}|k|^q\right)+
C_{22} N_{\frac{1}{2q}}\left(\frac{\sqrt{2}}{q}|k|^q\right)\right].
\end{equation}

Here $J_\nu(x)$ and $N_\nu(x)$ are Bessel functions and $I_\nu(x)$ and $K_\nu(x)$ are modified Bessel functions, see \cite{abr}. The asymptotics of $I_\nu(x)$ and $K_\nu(x)$ at $x \to \infty$
read \cite{abr}
\begin{equation}\label{asimp}
I_\nu(x)\approx \frac{e^x}{\sqrt{2\pi x}},\ K_\nu(x)\approx \sqrt{\frac{\pi}{2x}} e^{-x},
\end{equation}
while as $k \to -\infty$ the asymptotics of the functions $J_\nu(x)$ and $N_\nu(x)$ is oscillatory \cite{abr}.
This means that to  obtain a  localized pdf, we should leave the term with $K_{\frac{1}{2q}}$ in \eqref{terp4} only,
 so that $f(k)$ assumes following form
\begin{equation}\label{terp5}
f(k)=\left\{
\begin{array}{cc}
  C_{12}\sqrt{k}K_{\frac{1}{2q}}\left(\frac{\sqrt{2}}{q}k^q\right), & k \geq 0  \\ \\
  \sqrt{|k|}\left[C_{21} J_{\frac{1}{2q}}\left(\frac{\sqrt{2}}{q}|k|^q\right)+
C_{22} N_{\frac{1}{2q}}\left(\frac{\sqrt{2}}{q}|k|^q\right)\right], & k<0.   \\
   \end{array}
\right.
\end{equation}

As the equation \eqref{subs} is of second order, we should impose the continuity conditions  at $k=0$  for a  solution  and
 its first derivative.  We note, that the  numerical solution of equation \eqref{subs}  directly  involves the value of function
 and its first derivative at $k=0$.
That  implies (see Appendix \ref{sec:der} for  more  details)
\begin{equation}\label{terp6}
f(k)=C\sqrt{|k|}\left\{
\begin{array}{cc}
  K_\nu(u), & k \geq 0  \\ \\
  \frac{\pi}{2}\left[\cot \frac{\pi \nu}{2} J_\nu(u)-
 N_\nu (u)\right], & k<0,  \\
   \end{array}
\right.
\end{equation}
where $C \equiv C_{12}$,
\begin{equation}\label{terp6a}
\nu=\frac{1}{2q}\equiv \frac{1}{\mu+2},\ u=\frac{\sqrt{2}}{q}|k|^q\equiv \frac{2\sqrt{2}}{\mu+2}|k|^{1+\frac{\mu}{2}}.
\end{equation}
We note here that for $\mu=1$ we obtain from \eqref{terp6}
\begin{equation}\label{terp6b}
f(k)=C\sqrt{k}K_{\frac{1}{3}}\left(\frac{2\sqrt{2}}{3}k^{\frac 32}\right)=
C\frac{\pi \sqrt{3}}{2^{\frac 16}}{\rm {Ai}}\left(2^{\frac 13}\ k\right),
\end{equation}
known from Ref.~\cite{gar1}, where  the eigenvalue problem  for the Cauchy oscillator has been solved, see also Ref.~\cite{lorinczi1}.

\begin{figure}
\centerline{\includegraphics[width=0.9\columnwidth]{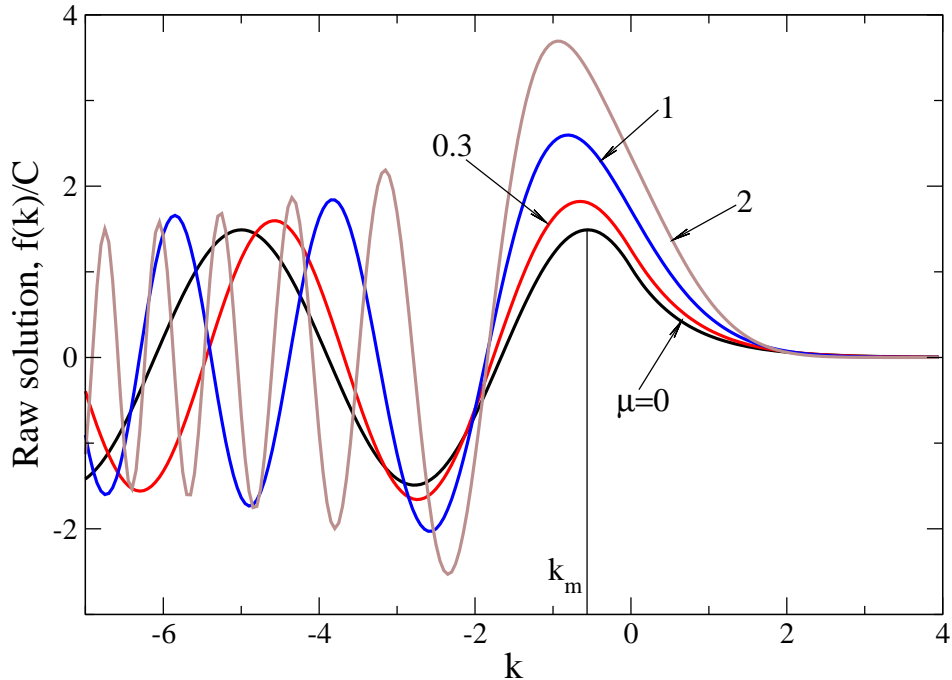}}
\caption{Raw solutions of Eq. \eqref{terp6}. Figures correspond to $\mu$ values. Solution for $\mu=1$
corresponds to Airy function \eqref{terp6b}. Formal solution for $\mu=0$ and the position of the first maximum of the oscillatory part $k_m$ are shown as an example.}
\label{f:raw}
\end{figure}

The next step is to find the position $k_m$ of the first maximum of an  oscillating part, next  shift the solution to the right by $k_m$,
 reflect the solution with respect to the  $y$ axis and "chop" the rest of oscillating parts.   By equating to zero the  first derivative
 of an  oscillating  contribution to \eqref{terp6}, after some algebra we get  the  following equation
\begin{equation}\label{max1}
N_{\nu-1}(u)-\cot\frac{\pi \nu}{2}J_{\nu-1}(u)=0,
\end{equation}
where $\nu$ and $u$ are defined by \eqref{terp6a}. Solutions of this equation can be tabulated, see  Table \ref{tab:max1}. The "raw" solutions \eqref{terp6} are shown (along with  the position of $k_m$) in Fig. \ref{f:raw}.

The normalization   condition allows us to fix the admissible values of hitherto unspecified constant $C$. Namely, we have
\begin{equation}\label{nomr}
C^2\int_{-\infty}^{\infty}f^2(k)dk= \ 2C^2 \int_{0}^{\infty}f^2(k)dk=2C^2\left[\int_{0}^{-k_m}f_1^2(k)dk+
\int_{-k_m}^{\infty}f_2^2(k)dk\right]=1,
\end{equation}
where $f_1$ and $f_2$ denote the oscillatory and decaying parts of Eq. \eqref{terp6} respectively. This integration can be performed numerically and results are reproduced in the right column of the Table \ref{tab:max1}.
The final step of the procedure is to invert the $k$-space solutions to $x$-space and square them to obtain the invariant  pdf.
Except for special $\mu $   cases, this procedure can be accomplished only numerically.

\begin{table}
\caption{Roots $k_m(\mu)$ of Eq.~\eqref{max1} corresponding to first maximum
of oscillatory part of  \eqref{terp6} for different $\mu$ (middle column) and
 normalization constants $C(\mu)$ (right column).} \label{tab:max1}
\begin{center}
\begin{tabular}{|c|c|c|}
\hline
$\mu$ &$k_m(\mu)$ & $C(\mu)$\\
\hline
0.0& -0.55536=$-\frac{\pi}{4\sqrt{2}} $ & 0.597135=$\sqrt{\frac{2}{\pi}}\left(1+\frac{\pi}{4}\right)^{-1/2}$\\
0.2& -0.621962 & 0.500134\\
0.4& -0.679458 & 0.429855\\
0.6& -0.729002 & 0.376894\\
0.8& -0.771717 & 0.335701\\
1.0& -0.808617 & 0.302823\\
1.2& -0.840577 & 0.276010\\
1.4& -0.868346 & 0.253745\\
1.6& -0.892550 & 0.234970\\
1.8& -0.913716 & 0.218927\\
2.0& -0.932286 & 0.205597\\
\hline
\end{tabular}
\end{center}
\end{table}

\section{Limiting (mis)behavior at the boundaries of $(0,2)\ni \mu $ and the zero (bottom) eigenvalue issue.}

The stability interval $\mu \in (0,2)$ is an open set. However, since $\mu$  can be chosen to be arbitrarily close, respectively to $0$ or $2$,
simply out of curiosity it is not useless to address  a hitherto unexplored  issue,  of what  is actually  going on   in the limiting  behaviors   of  $\mu \downarrow 0$ and $\mu \uparrow 2$, that can be consistently executed on the level of Fourier transforms.

We note that the operator $-|\Delta|^{\mu/2}$, as defined by Eq.~(\ref{def}),  is a pseudo-differential (Riesz) operator and the integral there-in needs to be taken as its Cauchy principal value. Hence both in the $x$ and $k$-spaces the point $0\in R$ is particularly  distinguished.

On formal grounds,  as exemplified in Section II, a naive expectation would amount to literal  setting of $\mu =0$ or $\mu =2$ instead of the "normal" stability index values $\mu \in (0,2)$. Then,  the    operator  $-|\Delta|^{\mu/2}$  would turn over into   $-|\Delta|^0\equiv -1$  and   $-|\Delta|\equiv \Delta $ respectively.
However, this compelling picture turns out to be problematic, except for rather special cases, like those considered before in Section II.

We recall that the  core of our solution method lies in  Fourier   transforming the compatibility condition (\ref{comp})
\begin{equation}\label{fourier}
\frac{x^2}{2}\rho ^{1/2}_*= - |\Delta |^{\mu/2}\rho ^{1/2}_* \rightarrow \frac{d^2 f(k)}{dk^2} =2|k|^\mu f(k)
\end{equation}
and actually solving the obtained  differential equation in the $k$-space.

Given the solution, the   unsettled problem is the validity of the inverse Fourier  transform   of the  outcome. That means that not only $f(k)$ itself, but also
$|k|^\mu f(k)$ and specifically $\frac{d^2 f(k)}{dk^2}$ need to be Fourier invertible to reproduce  the $x$-space version of the problem. One should as well be aware that with a resolved $\rho _*^{1/2}$, we  must have secured the  validity of the Fourier transform, that is explicit in  Eq.~\eqref{fourier}.

\subsection{Case of $\mu =0$}

We seriously address a Fourier transformed  equation  \eqref{fourier}  and
  employ again  Eqs.   \eqref{terp3} and  \eqref{subs}, while blindly   setting  $\mu =0$ there-in.
  The "raw" solution of \eqref{subs} can be either obtained from \eqref{terp5} or explicitly from \eqref{subs}. It has the form
\begin{equation}\label{fs1}
f(k)=\left\{
\begin{array}{cc}
  C_{12}e^{-k\sqrt{2}}, & k \geq 0  \\ \\
  C_{21}\cos k\sqrt{2}+C_{22} \sin k\sqrt{2}, & k<0.   \\
   \end{array}
\right.
\end{equation}
The continuity  condition at $k=0$ for  $f(k)$  reads
 \begin{equation}\label{fs2}
 C_{12}=C_{21}
 \end{equation}
 and for  the derivative
 \begin{equation}\label{fs3}
 C_{22}=-C_{12}=-C_{21},
 \end{equation}
 so that   (we set $C_{12}\equiv C$)
 \begin{equation}\label{fs4}
f(k)=C\left\{
\begin{array}{cc}
  e^{-k\sqrt{2}}, & k \geq 0  \\ \\
  \cos k\sqrt{2}- \sin k\sqrt{2}\equiv \sqrt{2}\cos\left(k\sqrt{2}+\frac{\pi}{4}\right), & k<0.   \\
   \end{array}
\right.
\end{equation}
The first maximum of oscillatory part is located  at
 \begin{equation}\label{fs5}
 k_m=-\frac{\pi}{4\sqrt{2}}\approx -0.555360367,
 \end{equation}
in accordance with Table \ref{tab:max1}. The "raw" solution \eqref{fs4} is shown (along with  the position of $k_m$) in
Fig. \ref{f:raw}.

Now we shift the whole solution to the right and "chop" the unnecessary piece of  an  oscillatory part
 \begin{equation}\label{fs6}
f(k)=C\left\{
\begin{array}{cc}
   \sqrt{2}\cos k\sqrt{2}, & 0 \leq k \leq -k_m  \\ \\
   e^{-(k+k_m)\sqrt{2}}, & k>-k_m.   \\
\end{array}
\right.
\end{equation}
The normalization condition  \eqref{nomr} reads
 \begin{equation}\label{fs7}
 2C^2\left[2\int_0^{-k_m}\cos^2 k\sqrt{2}dk+\int_{-k_m}^\infty e^{-2\sqrt{2}(k+k_m)}dk\right]=1
 \end{equation}
so that $C=\frac{1}{\sqrt[4]{2}\sqrt{1+\frac{\pi}{4}}}\approx 0.629325$.
This normalization coefficient is different from that in Table \ref{tab:max1}, because
the transition from Bessel functions of index 1/2 to elementary ones  introduces  an auxiliary  coefficient $\sqrt{\pi}/2^{3/4}$.
Thus, we have
 $C_{\text{table}}^2$=$(2\sqrt{2}/\pi)$$C^2$$=(2/\pi)/(1+\pi/4)$.
 This minor  difference  is insignificant with respect to the  normalizability of the  pertinent eigenfunction,  as an
 overall coefficient before $f(k)$ is just $C$, \eqref{fs7}.

Now we invert the Fourier transform to get
\begin{eqnarray}\label{fs8}
&&f(x)=\frac{2}{\sqrt{2\pi}}\int_0^\infty f(k)\cos \ kx \ dk=C\sqrt{\frac{2}{\pi}}\left[\sqrt{2}\int_0^{-k_m}\cos \ k\sqrt{2}\
\cos \ kx \ dk+ e^{-k_m\sqrt{2}}\int_{-k_m}^\infty e^{-k\sqrt{2}}\cos\ kx \ dk\right]= \nonumber \\
&&=C\sqrt{\frac{2}{\pi}}\frac{4}{x^4-4}\left(x\ \sin \frac{\pi x}{4\sqrt{2}}-\sqrt{2} \cos \frac{\pi x}{4\sqrt{2}}\right),\
C=\frac{1}{\sqrt[4]{2}\sqrt{1+\frac{\pi}{4}}}.
\end{eqnarray}

We note that  at the point $x=\pm \sqrt{2}$ function $f(x)$ looks divergent.  However, since the
terms in the numerator yield the same zero as in the denominator,   the  potentially dangerous  divergence is removed when one
approaches $x=\pm \sqrt{2}$.
The function $f(x)$ decays at  spatial  infinities as $1/x^3$ and is positive for $|x|<5$.
For  $|x|>5$  we encounter oscillations so that both zeroes and negative values are developed.
Thus,  it is clear the function $f(x)$ cannot
be interpreted  as an (arithmetic)  square root of a probability density  $\rho_*(x)$.  By this reason  alone  the obtained solution is  incongruent  with the $x$-space version of the compatibility condition.

More than that, an inversion of the Fourier transformed  equation  Eq. (\ref{fourier}) back to the x-space   leads to a contradiction, since  $\frac{x^2}{2} f(x) \neq  - f(x)$.
The roots of this discrepancy  are obvious. The  function (\ref{fs8})  at spatial  infinities  behaves as
$f(x)\sim {\frac{1}{x^3}} \sin \frac{\pi x}{4\sqrt{2}}$. Consequently, $x^2 f(x)\sim {\frac{1}{x}} \sin \frac{\pi x}{4\sqrt{2}}$ is not absolutely integrable on $R$  and thus not Fourier transformable.

\subsection{Case of $\mu =2$}
Concerning $\mu =2$, we recall   the eigenvalue  equation
  $(-\Delta + \frac{x^2}{2}-E_0)\rho ^{1/2}_*=0$, where $E_0=1/2$ is the lowest
eigenvalue for a quantum harmonic oscillator in the units $\hbar^2/2m=1$, $\omega ^2=m$, where $m$ is a particle  mass
 and $\omega$ is an oscillator  frequency. That gives rise to the Gaussian ground state function  as  a celebrated textbook result.

From this point of view the adopted (Section IV)  form of the compatibility condition (possibly, with a notable exception of
  $\mu =1$)  might seem  puzzling.
 Specifically,  the standard eigenvalue equation for the harmonic oscillator is plainly incompatible with   the
  $(-\Delta + \frac{x^2}{2})\rho ^{1/2}_*=0$  demand,  expected  (per force)  to correspond    to the case of $\mu =2$
   in the considerations of Section IV, c.f. Eq. {\ref{fourier}.

 Nonetheless,  the latter equation  seems  to    admit  a non-Gaussian  heavy-tailed  solution, which is derivable from
 \eqref{terp4}-\eqref{terp6}. In particular, an asymptotic
  ($k\rightarrow \infty $)  behavior of
\begin{equation}
K_{1/2q}\left({\frac{\sqrt{2}} q} k^q\right) \sim k^{-q/2}\,  \exp\left[-{\frac{\sqrt{2}}q} k^q\right]
\end{equation}
with $q=(\mu +2)/2$,  $\mu \in (0,2]$   yields an  asymptotics   in the $k$-space  for $\mu =2$, cf. \eqref{terp6},  as follows:
 $f(k) \sim   (1/\sqrt{k})  \, \exp[- k^2/\sqrt{2}]$. That  secures an absolute  convergence of the inverse Fourier    integral.

We have no sufficient analytic tools in hands to analyze various  features of the ultimate $x$-space solution. However, the performed  numerical (computer assisted) analysis proves that both $f(x)$ and $x^2f(x)$ are absolutely integrable. The  same occurs for $[x^2f(x)]^2$.
The pertinent $f(x) \equiv \rho ^{1/2}_*(x)$ has an inverse polynomial decay at infinities, decaying faster than $1/x^3$. In the vicinity of $x=0$  the obtained function mimics a Gaussian.  Those properties are visualized in Figs. 2 and 3.

The annoying point  that should be  clearly spelled out   in
connection with  ou  derivation procedure may pertain to the point
$0\in R$ and the fact that we have actually (albeit tacitly) defined
the Laplacian not on the whole of $R$, but on $\{R\backslash 0\}$. A
subsequent comment \cite{karw} indicates where the peculiarity of
our solution may possibly be rooted.
\vskip0.3cm
{\bf Comment 1:}  To make a clear distinction with the  text-book quantum mechanical reasoning  for the harmonic oscillator problem,  let us point out that we
have actually obtained appears to be related with the (normally discarded) "improper"  eigenvalue $n=-  1/2$
in the familiar  $E_n= \hbar \omega (n + 1/2)$   formula.
 Here we  emphasize  the
 text-book restriction upon the harmonic oscillator eigenvalues  $n\geq -1/2$, whose interpretation is: "all states
 corresponding to $n< -1/2$ need  to vanish identically", \cite{liboff}.  A  sufficient     condition to this end
is  that  the lowest energy  eigenfunction   vanishes under  the action of the   annihilation operator.  This actually selects $n=0$ as the bottom spectral
label  and $n\in (N\bigcup {0})$.
Our reasoning demonstrates that it is not a necessary condition, and actually,
 an "improper" eigenvalue $n=-1/2$ is admissible as a legitimate spectral label
  in the harmonic oscillator  problem.   However,  in this case  a specific choice of the boundary data (implicit in the
  construction procedure but hitherto  nor explicitly spelled out) appears to  enter the game. That may have an effect on the spectral features
  (including that of the proper domain definition and the   self-adjointness issue)  of the original problem, see  e.g.  the next Comment.

\vskip0.3cm {\bf Comment 2}, \cite{karw}: Generically, the Laplacian
$-\Delta $, if defined on a Schwartz space  $S(R)$  (fast decaying
functions, continuous  and
 with continuous derivatives),  is non-negative and essentially self-adjoint operator . Therefore its closure, which is a self-adjoint  operator, is
non-negative as well.  Its continuous spectrum extends from $0$ to $+\infty $.

However if we define the Laplacian on  $S(\{R\backslash 0\})$, the operator remains non-negative but ceases to be essentially self-adjoint. Its deficiency  indices  read $(2,2)$, implying an existence of a two-parameter family of self-adjoint extensions.
There is a well known in the literature example of such extensions determined by the boundary data $f'(+0)-f'(-0) = \alpha f(0)$.  On the operator level so point-wise  restricted   operator $-\Delta _{\alpha }$   may be  symbolically represented $-\Delta +  \alpha \delta (x)$, where $\delta (x)$  stands for  Dirac delta distribution.

For $\alpha =0$ we have a "normal" Laplacian. The case of $\alpha \rightarrow +\infty $  sets Dirichlet boundary condition at $0$ (an impenetrable barrier). If $\alpha >0$,  the operator $-\Delta _{\alpha }$  has a continuous non-negative absolutely continuous spectrum.

The case of $\alpha <0$ is particularly interesting, since then   $-\Delta _{\alpha }$ has one  non-degenerate isolated
negative eigenvalue  $-\alpha ^2/2$ with an eigenfunction $(|\alpha |^{1/2} \, \exp (-|\alpha x|/2)$, see e.g. \cite{viana}.
 The remaining part of the spectrum begins at $0$ and is absolutely continuous up to $+\infty $.

An operator $-\Delta   + x^2$,  defined on $S(\{R\backslash 0\})$, is regarded as positive and symmetric, but is not essentially self-adjoint. The deficiency indices are likewise $(2,2)$.  Assuming the previous boundary conditions we turn over to the one-parameter family of self-adjoint extensions $-\Delta _{\alpha } + x^2$, assuming likewise $\alpha <0$.

The choice of $\alpha =0$ leads to the standard harmonic oscillator problem. while for negative $\alpha $ there appears a fairly non-standard spectral solution described in detail in Ref. \cite{viana}. The problem has exactly one negative eigenvalue plus a positive discrete spectrum, of the form $\epsilon = (2\nu +1)/2 $.

Given negative $\alpha $, the   eigenvalues  $\epsilon$  come out from a transcendental equation
$\nu - \alpha \Gamma(1-\nu /2)/ \Gamma (1/2 - \nu /2)=0$.
   The value   $\epsilon =0$  appears for  $\nu = -1/2$ i.e. for
$\alpha =  -\Gamma (3/4)/ \Gamma (5/4)$.  The corresponding eigenfunction shows up  a $ \sim x^{-1/2} \exp(-x^2/2)$  decay at $+\infty $, \cite{viana}.

\begin{figure}
\centerline{\includegraphics[width=0.9\columnwidth]{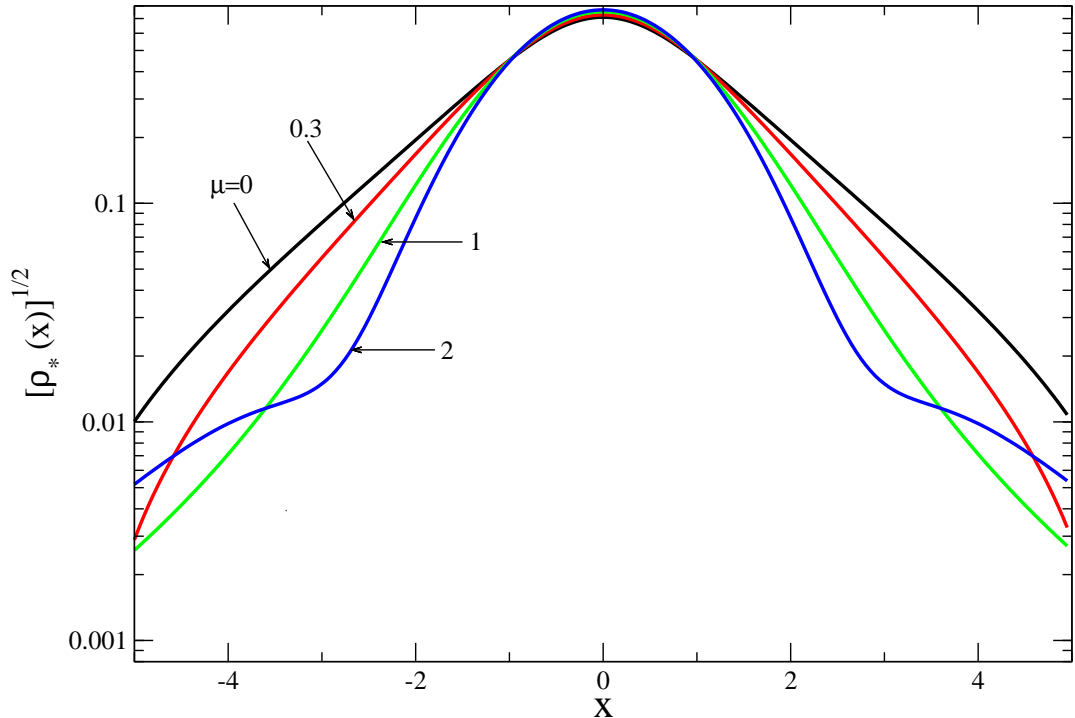}}
\caption{$f(x)=\rho ^{1/2}_*(x)$ for $\mu =0, 0.3, 1, 2$. The
logarithmic scale is employed to better visualize the behavior on
the "tails"} \label{f:sqrt}
\end{figure}

\begin{figure}
\centerline{\includegraphics[width=0.9\columnwidth]{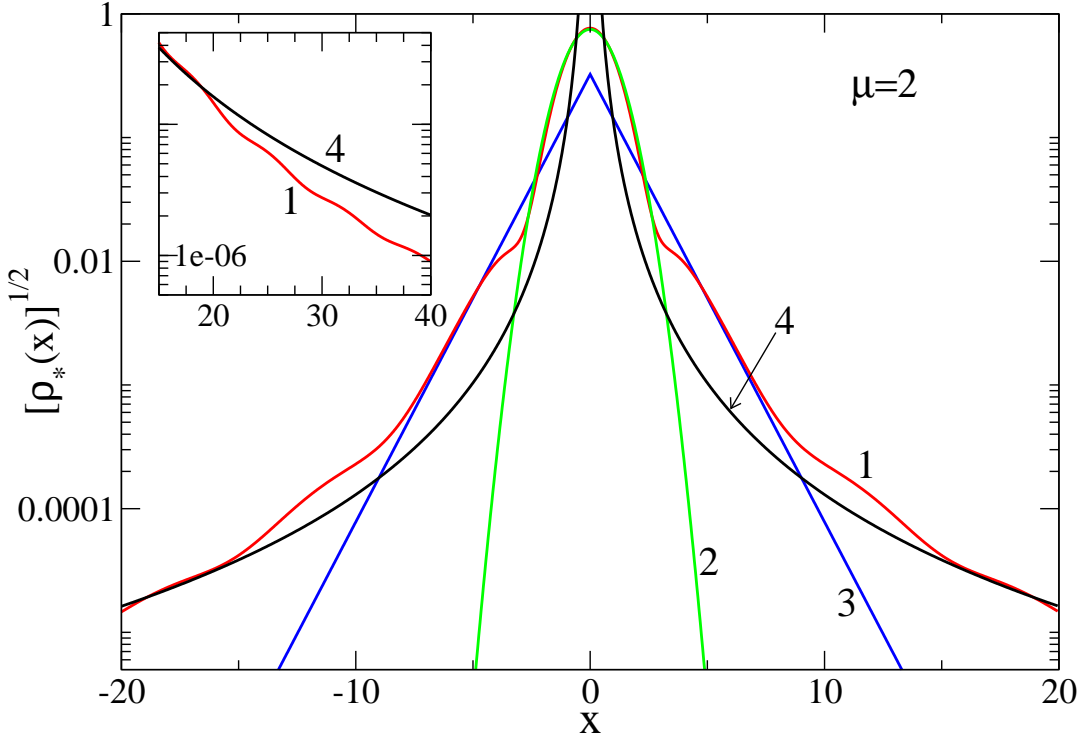}}
\caption{Different asymptotic fits (curves 2,3,4) of $f(x)$$=(\rho_*(x))^{1/2}$ (curve 1)
for $\mu=2$. Figures near curves:$ 2- 0.749\exp(-x^2/2), 3 - 0.15\exp(-|x|/1.2),
4 - 0.13/x^3$. Inset shows that $f(x)$ decays faster then $1/x^3$. }
\label{f:asimpt}
\end{figure}

\section{Conclusions}
To conclude, here we have presented   an outline of  the general formalism  for how  to find an  invariant pdf for a  pre-defined semigroup
 potential and  all stability index values $0<\mu<2$. We have  considered the  generic  case of symmetric even pdfs, for which  a
 potential ${\cal V}(x)$ is  an  even function.   That was dictated by known spectral properties of fractional
 operators associated with a symmetric stable noise.  We  have  reduced the problem of
finding a  terminal pdf  to that of  finding a  solution of the ordinary differential equation with infinite number of terms in momentum
space. Such differential equation, even if hard to handle analytically, can rather easily  be solved with a  numerical assistance.
For polynomial potentials the number of terms becomes finite and the pertinent  equation can be solved analytically.

The outlined procedure has been explicitly performed for the family of L\'{e}vy stable oscillators, with a common
 quadratic semigroup potential ${\cal V}(x)=x^2/2$.   In this case,  a solution of  the  corresponding differential equation
 in  the  $k$-space has been obtained by employing a suitable  continuity  procedure at $k=0$.
 After Fourier-inversion and squaring the result, this  yields an ultimate functional  form
of the desired   invariant (and possibly  terminal)  pdf  of the jump-type process,  for arbitrary $\mu \in (0,2)$.

 We  have  analyzed a limiting behavior of solutions in the vicinity and at the boundaries $\mu=0$ and $2$  of an open stability interval $(0,2)$.
  In the case of $\mu=0$  we have derived an explicit analytical form of the solution in $k$ and (by Fourier inversion) in  the $x$-spaces. The pertinent function shows a nontrivial oscillating behavior and thus is definitely not a square root of any pdf.  However,  with  an explicit analytic form of $f(x)$ one easily  proves that the
$x^2f(x)$ is not-Fourier transformable, hence the crucial connection between the x-space and k-space solutions, Eq. (\ref{fourier}) has been lost.

We have  shown  that for  $\mu=2$  a positive $\rho _*^{1/2}(x)$ is obtained.  This  invariant  solution  definitely  corresponds
to the  "improper eigenvalue" $\epsilon =0$   of the energy operator. To comply with the common knowledge about the spectral
 properties of the operator  $-\Delta +x^2$,  we find that  the (hitherto neglected) presence of the special boundary data
 has been enforced. Then only the  self-adjoint extensions of the  corresponding non-negative  operator may be introduced,
  and  $\rho ^{1/2}_*(x)$ may receive a consistent interpretation as  an asymptotic (terminal) target in the semigroup evolution.

 This particular case in turn, indicates the main jeopardy hidden in the presented reconstruction problem. Once an invariant pdf has been settled,
  it is necessary to identify a  consistent self-adjointness domain for the involved operator $\hat{H}$. Then only an invariant pdf may be interpreted as an asymptotic
  (terminal) one for a well defined semigroup  $\exp(-t  \hat{H} )$.

   A problem of reconstructing a self-adjoint  operator from the knowledge of its ground state is not trivial at all, see e.g. for an investigation of \cite{vilela} in the
   diffusion-type framework. In the harmonic oscillator problem of Comment 2 it is known that a corresponding self-adjoint Hamiltonian shares
   odd-parity (label $n=1,3,5,...$)  part of the spectrum with a "normal" harmonic oscillator, while an
    even-labeled part is substantially different (that pertains to the associated eigenfunctions as well).\\

{\bf Acknowledgement: }
We are willing words of gratitude to Professor Witold Karwowski for enlightening discussion  and explanation of  the role of boundary data in
differentiating between the "normal"
and "abnormal" (zero bottom eigenvalue) harmonic oscillator  spectral  problems.

\appendix

\section{Continuity  conditions at $k=0$}\label{sec:der}

From \eqref{terp5}, the functions at $k=0$ read
\begin{equation}\label{join1}
C_{12}[\sqrt{k}K_\nu(u)]_{k=0}=C_{22}[\sqrt{k}N_\nu(u)]_{k=0}.
\end{equation}
The derivatives at $k=0$
\begin{equation}\label{join2}
C_{21}[\sqrt{k}J_\nu(u)]'_{k=0}+C_{22}[\sqrt{k}N_\nu(u)]'_{k=0}=C_{12}[\sqrt{k}K_\nu(u)]'_{k=0}.
\end{equation}
Such forms of \eqref{join1} and \eqref{join2} are dictated by the following asymptotic expansions of Bessel functions
near $k=0$ in variables \eqref{terp6a}
\begin{eqnarray}
&&K_\nu(u)\approx \frac{\Gamma(-\nu)}{2}\sqrt{k}\left[\frac{\sqrt{2}}{\mu+2}\right]^\nu+\frac{\Gamma(\nu)}{2\sqrt{k}}\left[\frac{\mu+2}{\sqrt{2}}\right]^\nu,\nonumber \\
&&N_\nu(u)\approx -\frac{\cos \pi \nu \ \Gamma(-\nu)}{\pi}\sqrt{k}\left[\frac{\sqrt{2}}{\mu+2}\right]^\nu-\frac{\Gamma(\nu)}{\pi\sqrt{k}}\left[\frac{\mu+2}{\sqrt{2}}\right]^\nu,\nonumber \\
&&J_\nu(u)\approx \frac{\sqrt{k}}{\Gamma(1+\nu)} \left[\frac{\sqrt{2}}{\mu+2}\right]^\nu.\label{join3}
\end{eqnarray}
Eq. \eqref{join3} means that $[\sqrt{k}J_\nu(u)]_{k=0}=0$. For reference purposes, the derivatives like $[\sqrt{k}N_\nu(u)]'_{k=0}$ (we first multiply by $\sqrt{k}$ and then differentiate) read
\begin{eqnarray}
&&[\sqrt{k}N_\nu(u)]'_{k=0}=-\frac{\cos \pi \nu \ \Gamma(-\nu)}{\pi}\left[\frac{\sqrt{2}}{\mu+2}\right]^\nu , \
[\sqrt{k}K_\nu(u)]'_{k=0}= \frac{\Gamma(-\nu)}{2}\left[\frac{\sqrt{2}}{\mu+2}\right]^\nu, \nonumber \\
&&[\sqrt{k}J_\nu(u)]'_{k=0}=\frac{1}{\Gamma (1+\nu)}\left[\frac{\sqrt{2}}{\mu+2}\right]^\nu. \label{join4}
\end{eqnarray}
Substitution of values of functions and derivatives into \eqref{join1} and \eqref{join2} yields
\begin{equation}\label{join5}
C_{22}=-\frac{\pi}{2}C_{12},\ C_{21}=\frac{\pi}{2}C_{12}\cot \frac{\pi \nu}{2},
\end{equation}
which, after employing the identity $ \Gamma(1+\mu)\Gamma(-\mu)=- \pi /\sin (\pi \mu )$
 gives rise to  Eq.~\eqref{terp6}.

\end{document}